\documentclass[rmp,nofootinbib,onecolumn,floatfix]{revtex4}
\usepackage{graphicx}
\bibpunct[, ]{[}{]}{,}{a}{,}{,}
\ifnum\lefthyphenmin<2\lefthyphenmin=2\fi
\ifnum\righthyphenmin<2\righthyphenmin=2\fi

\begin{document}
\title{Information Optimization in Coupled Audio--Visual Cortical Maps}
\author{Mehran Kardar}
\affiliation{Department of Physics, Massachusetts Institute of Technology,
Cambridge, Massachusetts 02139}
\author{A. Zee}
\affiliation{Kavli Institute for Theoretical Physics, University of California,
Santa Barbara, California 93106}
\date{\today }
\begin{abstract}
Barn owls hunt in the dark by using cues from both sight and sound to locate their prey. This task is facilitated by topographic maps of the external space formed by neurons (e.g., in the optic tectum) that respond to visual or aural signals from a specific direction. Plasticity of these maps has been studied in owls forced to wear prismatic spectacles that shift their visual field. Adaptive behavior in young owls is accompanied by a compensating shift in the response of (mapped) neurons to auditory signals. We model the receptive fields of such neurons by linear filters that sample correlated
audio--visual signals, and search for filters that maximize the gathered information, while subject to the costs of rewiring neurons. Assuming a higher fidelity of visual information, we find that the corresponding receptive fields are robust and unchanged by artificial shifts. The shape of the aural receptive field, however, is controlled by correlations between sight and sound. In response to prismatic glasses, the aural receptive fields shift in the compensating direction, although their shape is modified due to the costs of rewiring.
\end{abstract}
\maketitle
%\preprint{\tighten \vbox{\hbox{hep-th/0208029} }}
%\pacs{}
%\newpage
\section{Introduction}

In the struggle of biological organisms to survive and reproduce,
processing of information is of central importance. 
Sensory signals provide valuable information about the external world,
such as the locations of predators and preys.
Localization of sources is facilitated by topographic maps of neurons to various
parts of the brain\cite{maps}, reflecting the spatial arrangements of signals around the animal.
The barn owl has to rely extensively on sounds to find its prey in the dark,
and has consequently developed precise `auditory space maps.'

By extensive experiments, Knudsen and collaborators have shown that
the optic tectum of the barn owl has both visual and aural maps of space
that are in close registry\cite{K82,K83}.
The visual signal plays a crucial role in aligning the aural map; experimental
manipulations of the owl's sensory experience reveal the plasticity of these
maps in young animals, and the instructive role played by the visual experience.
(A recent review, with specific references can be found in Ref.~\cite{Krev}.)
The current study was motivated by experiments in which owls are fitted 
with prismatic spectacles that shift the visual fields by a preset degree in 
the horizontal direction\cite{BK}.
In young owls, the receptive auditory maps were found to shift to remain in registry
with the visual maps, which stayed unchanged.

There is at least one theoretical attempt to explain the registry of neural maps
through a `value--dependent learning,' where synaptic connections in a network are enhanced
after `foveation towards an auditory stimulus'\cite{Rucci}.
In this paper we take a more abstract approach to the coupling of audio--visual maps, 
and search for neural connections (receptive fields) that maximize the 
information gained from the sensory signals.
In earlier studies\cite{BRZ,BZ},  Bialek and one of us formulated an approach to
optimization of information in the visual system, and in computations with
neural spike trains\cite{spikes}.

Here, we extend the methods of Ref.~\cite{BRZ} for computing receptive fields
in the visual system, to finding the optimal connectivities in an audio-visual cortex,
such as the owl's optic tectum.
We find that the shape and registry of the aural map is established by the correlations
between the audio and visual signals.
In response to an artificial shift of the visual field (as with the prismatic spectacles),
the visual receptive field is unchanged.
While the aural receptive field shifts in the adaptive direction, its shape changes due to the
costs of rewiring the neurons.

The general formalism for our calculations is set up in Sec.~\ref{formalism},
which reviews the methodology introduced in Ref.~\cite{BRZ}.
The essence of this approach is the assumption that neural connections act as
{\em linear} filters of the incoming signals, and also introduce noise in the outputs.
If the (correlated) input signals, and the random noise, are taken from Gaussian probability
distributions, the outputs are also Gaussian distributed.
The Shannon\cite{information} information content of the resulting outputs is easily calculated.
The task is to find filter functions that maximize this information, subject
to  biologically motivated costs, and for given correlations of the input signals.
In Ref.~\cite{BRZ} this approach was used to obtain receptive fields in the visual system.
In Sec.~\ref{AVinputs}, we generalize this formalism to coupled audio-visual signals.

A necessary input to the calculations is the correlations between the audio and visual 
signals, as discussed in Sec.~\ref{S-correlations}.
Since it is clearly much easier to localize objects by sight that sound, it is reasonable
that the information carried by the visual channel should far exceed the aural one.
The two sources of information are however quite likely to be correlated, 
resulting in couplings between the corresponding filters.
In the experiments on barn owls, the prismatic glasses shift the visual field and
hence modify the correlations between the signals. 
We examine how such shifts change the filter functions (neural connectivities) that
optimize the information content in the outputs.

As argued in Sec.~\ref{results}, the disparities in the strengths of visual and aural 
signals simplify the search for optimal filters.
In particular, we find that the visual receptive fields are relatively robust and unchangeable,
while the shape of the aural receptive field is the product of two terms:
One reflects the correlations between sights and sounds, and shifts along with
external displacement of these signals; 
the second is associated with the costs of making connections to distant neurons.
This result is further  interpreted in the final section (Sec.~\ref{discussions}),
where some implications for experiments, as well as directions for future extensions
and generalizations, are also discussed.

\section{Analysis of Information}\label{analysis}

\subsection{General Formalism}\label{formalism}

The processing of information by neural connections in the cortex is
modelled in Ref.~\cite{BRZ} as follows:
After passing through intermediate stations, sensory signals arrive as
a set of inputs $\{s_{J}\}$.
Further processing takes place by neurons that sample the information from
a subset of these inputs, and produce an appropriate output.
For ease of calculation, the outputs are represented as a linear transformation
of the inputs, according to
\begin{equation}
\label{filter}
O_{i}=\sum_J F_{iJ}s_{J}+\eta _{i}\quad.
\end{equation}
The filtering of information is thus parameterized by the matrix $\{F_{iJ}\}$,
and is also assumed to introduce an unavoidable  noise $\eta _{i}$.
There are of course many possible sensory inputs, which can be taken from
a joint probability distribution $P_{in}[s_{J}]$.
Equation~(\ref{filter}) is thus  a transformation from one set
of random variables (the inputs) to another (the outputs); the latter
described by the joint probability distribution function $P_{out}[O_{i}]$.
The amount of information associated with a given probability distribution
is quantified\cite{information} (up to a baseline and units) by 
${\cal I}[P]\equiv-\left\langle\ln~P\right\rangle$, where the averages are 
taken with the corresponding probability.
The task of finding optimal filters is thus to come up with the matrix $F$ that
maximizes ${\cal I}[P_{out}]$ for specified input and noise probabilities.

The Shannon information can be calculated easily for Gaussian distributed
random variables.
Let us consider the set of $N$ random variables $\{x_i\}$, taken from the probability
\begin{equation}
\label{gaussian}
P\left[x_i\right]=\sqrt{\frac{\det A}{(2\pi)^N}}\exp\left[-\frac{1}{2}x_i A_{ij}x_j\right],
\end{equation}
where summation over the repeated indices is implicit, and $\det A$ indicates
the determinant of the $N\times N$ matrix with elements $A_{ij}$.
It is easy to check that, up to an unimportant additive constant of $N/2$, 
\begin{equation}
\label{gauss-info}
{\cal I}[P]=-\frac{1}{2}\ln\det A=\frac{1}{2}\ln\det\left[\left\langle x_i x_j\right\rangle\right],
\end{equation}
where we have noted that the pairwise averages are related to the inverse
matrix by $\left\langle x_i x_j\right\rangle=A_{ij}^{-1}$.
A {\em linear filter} as in Eq.~(\ref{filter}), maps one set of Gaussian variables to new ones. 
Thus if we assume that the inputs $\{s_J\}$, and the (independent) noise $\{\eta_i\}$,
are Gaussian distributed, we can calculate the information content of the 
output using Eq.~(\ref{gauss-info}), with
\begin{equation}
\label{gauss-corr}
\left\langle O_i O_j\right\rangle=F_{iJ}F_{kL}
\left\langle s_J s_L\right\rangle+\left\langle \eta_i \eta_j\right\rangle.
\end{equation}

We are interested in describing cortical maps related to visual or aural
localization of objects.
These locations vary continuously in space, and are topographically mapped
to positions on a two-dimensional cortex.
As such, it is convenient to promote the indices $i$ and $J$, used above to label output
and input neurons, to continuous vectors in two dimensional space.
For example, following Ref.~\cite{BRZ}, let us consider 
an image described by a scalar field $s(\vec{x}\,)$ on a $2-$dimensional surface
with coordinates $\vec{x}.$ The image is sampled by an array of cells such
that the output of the cell located at $\vec{x}$ is given by
\begin{equation}
\label{inout}
O\left(\vec x\,\right)=\int d^{2}yF\left(\vec{x}-\vec{y}\,\right)s\left(\vec{y}\,\right)
+\eta\left(\vec x\,\right)  ,
\end{equation}
where the function $F(\vec r\,)$ describes the {\em receptive field} of the cell.
Assuming uncorrelated neural {\em noise}, 
$\langle \eta(\vec x\,)\eta(\vec x^{\,\prime})\rangle=N\delta^2(\vec x-\vec x^{\,\prime})$,
and {\em signal} correlations 
$\langle s(\vec x\,)s(\vec x^{\,\prime})\rangle=S(\vec x-\vec x^{\,\prime})$,
the filter-dependent part of the output information is given by
\begin{equation}
\label{info}
{\cal I}=\frac{1}{2}\ln\det\left[\delta^2 (\vec x-\vec x^{\,\prime})+\int d^{2}y\int d^{2}y^{\,\prime}
F\left(\vec{x}-\vec{y}\,\right)F\left(\vec{x}^{\,\prime}-\vec{y}^{\,\prime}\,\right)
\frac{S\left(\vec{y}-\vec{y}^{\,\prime}\,\right)}{N}\right].
\end{equation}
Note that we have assumed that the signal is {\em translationally invariant},
such that correlations only depend on the relative distance between their sources.
This allows us to change basis to the Fourier components, 
$\tilde s(\vec{k}\,)\equiv\int d^2x \exp(-i\vec{k}\cdot\vec{x}\,)s(\vec{x}\,)$,
which are uncorrelated for different wave-vectors $\vec{k}$.
The overall information is then obtained from a sum of independent contributions,
and using $\sum_{\vec k}\to A\int d^2k/(2\pi)^2$ where $A$ is the cortical area,
equal to
\begin{equation}
\label{infoK}
{\cal I}=\frac{A}{2}\int\frac{d^2k}{(2\pi)^2}
\ln\left[1+\left|F(\vec{k}\,)  \right|^2 {\cal S}(\vec{k}\,)  \right],
\end{equation}
where $F(\vec{k}\,)$ and $\tilde{S}(\vec{k}\,)$ are Fourier transforms of
the receptive field $F(\vec{x}\,)$, and signal to noise correlations
$S(\vec{x}\,)/N$, respectively.
 
The task is to find the function $F(\vec{k}\,)$ which maximizes the information ${\cal I}$.
Clearly, we need to impose certain costs on this function, since otherwise
the information gain can become enormous for $F\to\infty$.
This cost ultimately originates from the difficulties of creating and maintaining
neural connections that gather and transmit information over some distance,
and is hard to quantify.
Following Ref.~\cite{BRZ}, we shall assume that the overall cost (in appropriate
`information' units) has the form
\begin{equation}
\label{cost}
{\cal C}=\int d^2x C\left(\vec{x}\,\right)F\left(\vec{x}\,\right)^2\approx
\int d^2x\frac{\lambda +\mu x^2}{2}F\left(\vec{x}\,\right)^2=
\frac{A}{2}\int\frac{d^2k}{(2\pi)^2}\left[\lambda\left|F(\vec{k}\,)  \right|^2
+\mu\left|\vec\nabla_{k}F(\vec{k}\,)  \right|^2\right].
\end{equation}
This expression can be regarded as an expansion in powers of $F$ and
$x$, with the assumption that the cost is invariant under changing the
sign of $F$, and independent of the direction of the vector $\vec{x}$.
It imposes a penalty for creating connections which increases quadratically
with the length of the connection.
Our central conclusion is in fact insensitive to the form of $C(\vec x\,)$.

If the costs are prohibitive, there will be no filtering of signals. 
To avoid such cases, we compare only filters that are constrained such
that $\int d^2x F\left(\vec{x}\,\right)^2=1$ (or any other constant).
In the optimization process, this constraint can be implemented via a Lagrange
multiplier, resulting in an effective cost similiar to the term proportional to
$\lambda$ in Eq.~(\ref{cost}). Thus, this term and the constraint can be
used interchangably.
In Ref.~\cite{BRZ} it was shown that optimizing Eq.~(\ref{infoK}) subject
to the cost of Eq.~(\ref{cost}), in the limit of low signal to noise, is
equivalent to solving a Schr\"odinger equation
%\begin{equation}\label{Sch}
%\left[\mu\nabla_k^2 F+{\cal S}(k)\right]F(\vec{k}\,)  =\lambda F(\vec{k}\,)  .
%\end{equation}
with $F(\vec{k}\,)$ playing the role of the wave function in a potential ${\cal S}(k)$,
and the Lagrange multiplier taking the value of the ground state energy.
A potential of the form ${\cal S}(k)\propto k^{-2}$ was there used to obtain
receptive fields with on center/off surround character.
In the next section we generalize this approach by considering correlated visual
and aural inputs.

\subsection{Coupled Audio--Visual Inputs}\label{AVinputs}

In our idealized model, a neuron in the optic  tectum of the owl filters input
signals coming from both the visual and auditory systems, and its output is
given by the generalization of Eq.~(\ref{inout}) to
\begin{equation}\label{AVout}
O\left(\vec{x}\,\right)=\int d^{2}yF_{\alpha}\left(\vec{x}-\vec{y}\,\right)
s_\alpha\left(\vec{y}\,\right)+\eta\left(\vec{x}\,\right),
\end{equation}
where $\alpha$ is summed over A and V for audio and visual signals, respectively.
Assuming as before that the signals $s_\alpha$ and the noise $\eta$ are independent,
correlations of the output are obtained as
\begin{equation}
\label{av-corr}
\left\langle O\left(\vec{x}_{1})O(\vec{x}_{2}\right)\right\rangle =
\int d^{2}y_{1}\int d^{2}y_{2}F_\alpha\left(\vec{x}_{1}-\vec{y}_{1}\right)
F_\beta\left(\vec{x}_{2}-\vec{y}_{2}\right)S_{\alpha\beta}
\left(\vec{y}_{1}-\vec{y}_{2}\right)+N\delta^2 \left(\vec{x}_{1}-\vec{x}_{2}\right).
\end{equation}
For translationally invariant signals, the output information is given by the generalization
of Eq.~(\ref{infoK}) to
\begin{equation}
\label{av-info}
{\cal I}=\frac{A}{2}\int\frac{d^2k}{(2\pi)^2}
\ln\left[1+F_\alpha(\vec{k}\,)   {\cal S}_{\alpha\beta}(\vec{k}\,)  
F_\beta(-\vec{k}\,)\right],
\end{equation}
where ${\cal S}_{\alpha\beta}(\vec{k}\,)$ is a $2\times 2$ matrix of (Fourier transformed)
signal to noise correlations.

Once more, we have to impose some constraints in order to make the maximization
of the information in Eq.~(\ref{av-info}) with respect to the functions $F_{\rm V}$ and $F_{\rm A}$
biologically meaningful. 
In principle, there could be different costs for connections processing aural and visual
signals. In the absence of concrete data, we make the simple choice of using the
same form as Eq.~(\ref{cost}) for both sets of filters, so that the overall cost is
\begin{equation}
\label{av-cost}
{\cal C}=\frac{A}{2}\int\frac{d^2k}{(2\pi)^2}\left[\lambda 
F_\alpha(\vec{k}\,)  F_\alpha(-\vec{k}\,)  +\mu
\vec\nabla_{k}F_\alpha(\vec{k}\,)  \cdot \vec\nabla_{k}F_\alpha(-\vec{k}\,)  \right].
\end{equation}
The first term in the above cost function can again be interpreted as
a Lagrange multiplier $\lambda$ imposing a normalization constraint
\begin{equation}
\label{norm}
\int d^2x F_\alpha(\vec x\,)^2=A \int \frac{d^2k}{(2\pi)^2}
F_\alpha(\vec{k}\,)  F_\alpha(-\vec{k}\,)=1.
\end{equation}

\subsection{Signal Correlations}\label{S-correlations}
To proceed further, we need the matrix of signal to noise correlations, which has the form
\begin{eqnarray}
{\cal S}(\vec k\,)=\begin{pmatrix}
{  {\cal S}_{\rm V}(k)&{\cal R}(k)e^{i\vec k\cdot\vec c}\nonumber\\ 
{\cal R}(k)e^{-i\vec k\cdot\vec c}&{\cal S}_{\rm A}(k)} 
\end{pmatrix}.
\end{eqnarray}
%\begin{itemize}
%\item 
The diagonal terms represent the self correlations of each signal. 
%In Ref.~\cite{BRZ}, these correlations were assumed to have a scale invariant form,
%${\cal S}_\alpha(k)\propto 1/k^2$.
%To ease calculations, and to explore the role of different scales, we shall assume
%correlations of Gaussian form 
%\begin{equation}\label{S-gauss}
%{\cal S}_\alpha(k)=\frac{{\bf S}_\alpha}{2\pi l_\alpha}\exp\left(-\frac{k^2l_\alpha^2}{2}\right),
%\quad{\rm for}~\alpha={\rm V~or~A},
%\end{equation}
%where ${\bf S}_\alpha$ sets the overall strength of the signal, and $l_\alpha$ is an (angular)
%range over which signals are likely to be correlated.
%(As we shall demonstrate, the actual shapes are not so important.)
% \item 
Since many sources generate both sight and sound, the audio and visual signals
will be correlated. These correlations are captured by the off-diagonal term ${\cal R}(k)$. 
%which we shall assume to be a Gaussian of strength ${\bf S}_{\rm X}$, 
%and characteristic scale $l_{\rm X}$.
In the experiments on owls\cite{BK}, the visual signal is artificially displaced by a 
fixed angle in the horizontal direction. If we indicate this angle by the vector $\vec c$,
an aural signal at location $\vec x$ becomes correlated with a visual signal at
$(\vec x+\vec c\,)$. After Fourier transformation, this shift appears as the exponential
factor $\exp(i\vec k\cdot\vec c\,)$ in the off-diagonal terms of the correlation matrix.
% \item 

So far, we have treated sight and sound on the same footing. 
It is reasonable to assume that under most (well lit) conditions the quality of visual
information is much higher than the aural one.
For ease of computation, we shall further assume that the actual signal to noise ratio
is quite small, resulting in the set of inequalities
\begin{equation}
\label{inequalities}
{\cal S}_{\rm A}(\vec{k}\,)\ll {\cal R}(\vec{k}\,)\ll {\cal S}_{\rm V}(\vec{k}\,)\ll 1.
\end{equation}
%\end{itemize}
In this limit of small signal to noise, the logarithm in Eq.~(\ref{av-info}) can be approximated
by its argument (without the one), resulting in a quadratic form in the filter functions.
Our task then comes down to maximizing the function
\begin{eqnarray}
\label{W}
{\cal W}\left[F_\alpha(\vec{k}\,)\right] \equiv{\cal I}-{\cal C} &=&
\frac{A}{2}\int \frac{d^{2}k}{(2\pi )^{2}}\left[
{\cal S}_{\rm V}(\vec{k}\,)\left|F_{\rm V}(\vec{k}\,)\right|^{2}+
{\cal S}_{\rm A}(\vec{k}\,)\left|F_{\rm A}(\vec{k}\,)\right|^{2} \right. \nonumber \\
&&+{\cal R}(\vec{k}\,)\left(F_{\rm V}(\vec{k}\,)F_{\rm A}^{*}(\vec{k}\,)
e^{-i\vec{k}\cdot \vec{c}}+
F_{\rm V}^{*}(\vec{k}\,)F_{\rm A}(\vec{k}\,)e^{i\vec{k}\cdot \vec{c}}\right) \nonumber \\
&&\left.
-\lambda \left(\left|F_{\rm V}(\vec{k}\,)\right|^{2}+
\left|F_{\rm A}(\vec{k}\,)\right|^{2}\right)
-\mu \left(\left|\vec{\nabla}_{k}F_{\rm V}(\vec{k}\,)\right|^{2}+
\left|\vec{\nabla}_{k}F_{\rm A}(\vec{k}\,)\right|^{2}\right)
\right],
\end{eqnarray}
with respect to $F_{\rm V}$ and $F_{\rm A}$.

\subsection{Results}\label{results}

The optimal filters are obtained from functional derivatives of Eq.~(\ref{W}).
Setting the variations with respect to $F_{\rm V}^{*}(\vec{k}\,)$ to zero gives
\begin{equation}\label{varV}
\left[{\cal S}_{\rm V}(\vec{k}\,)+\mu\vec{\nabla}_{k}^{2}\right]F_{\rm V}(\vec{k}\,)
=\lambda F_{\rm V}(\vec{k}\,)
%+{\cal O}\left({\cal R}F_{\rm A}\right),
-\left\{{\cal R}(\vec{k}\,)F_{\rm A}(\vec{k}\,) e^{i\vec{k}\cdot \vec{c}}\right\},
\end{equation}
while $\delta W/\delta F_{\rm A}^{*}=0$, leads to
\begin{equation}\label{varA}
\left[\lambda-\mu\vec{\nabla}_{k}^{2}\right]F_{\rm A}(\vec{k}\,)
={\cal R}(\vec{k}\,)F_{\rm V}(\vec{k}\,) e^{-i\vec{k}\cdot \vec{c}}+
%{\cal O}\left({\cal S}_{\rm A}F_{\rm A}\right).
\left\{{\cal S}_{\rm A}(\vec{k}\,)F_{\rm A}(\vec{k}\,)\right\}.
\end{equation}
In arranging the above equations, we have placed within curly brackets terms that
are much smaller according to the hierarchy of inequalities in Eq.~(\ref{inequalities}).
Note that in the absence of any correlations between the two signals (${\cal R}=0$),
$F_{\rm A}=0$, since the aural signal is assumed to be much weaker than the
visual one. Any non-zero $F_{\rm A}$ reduces $F_{\rm V}$ due to the normalization
condition, resulting in a smaller value of ${\cal W}$.
It is indeed the correlations between the two signals that lead to a finite value of
$F_{\rm A}$, of the order of $({\cal R}/{\cal S}_{\rm V})$.
(Since $\lambda\sim{\cal O}({\cal S}_{\rm V})$ as shown below.)

To leading order, Eq.~(\ref{varV}) is the Schr\"odinger equation obtained in Ref.~\cite{BRZ}
for the visual receptive field. Without further discussion, we shall indicate its
solution by
\begin{equation}\label{Fv}
F_{\rm V}(\vec{x}\,)=F^0_{\rm V}(\vec{x}\,),\quad{\rm and}\quad 
\lambda=E_{\rm V}={\cal O}\left({\cal S}_{\rm V}\right).
\end{equation}
Note that we don't imply that cells in the optic tectum should have a receptive field
for visual signals identical to that in the visual cortex.
The quality of signals, the costs of neural connections, and the
response of the cells may well vary from one cortical area to another.
The eigenvalue $E_{\rm V}$ is controlled by the strength of the visual correlations
and is of the order of ${\cal S}_{\rm V}$.

%The solution to Eq.~(\ref{varA}) can be written symbolically as
%\begin{equation}\label{Fak}
%F_{\rm A}(\vec{k}\,)=\left[E_{\rm V}-\mu\vec{\nabla}_{k}^{2}\right]^{-1}
%\left[{\cal R}(\vec{k}\,)\Psi_{\rm V}(\vec{k}\,) e^{-i\vec{k}\cdot \vec{c}}\right].
%\end{equation}

To simplify the solution to Eq.~(\ref{varA}), we first assume that ${\cal R}(\vec{k}\,)=R$,
a constant independent of $\vec k$. 
This is quite a reasonable assumption, corresponding to visual and aural signals
that are correlated only if coming from the same direction, i.e. with
$\langle s_{\rm V}(\vec x_1)s_{\rm A}(\vec x_2)\rangle=R\delta^2(\vec x_1-\vec x_2)$.
We can then Fourier transform the two sides of this equation to obtain
\begin{equation}\label{varAx}
\left(E_{\rm V}+\mu x^2\right)F_{\rm A}(\vec x\,)=RF^0_{\rm V}(\vec{x}-\vec{c}\,),
\end{equation}
and quite generally, for an arbitrary form of the cost function in Eq.~(\ref{cost}),
the solution is
\begin{equation}\label{Fa}
F_{\rm A}(\vec x\,)=\frac{R}{E_{\rm V}+C_A(x)} F^0_{\rm V}(\vec{x}-\vec{c}\,).
\end{equation}
Due to the quadratic form of Eq.~(\ref{W}), the above result is the linear
response of the system to the correlations between signals.

The significance of our result is that the aural receptive field $F_{A}(\vec{x}\,)$ 
is not simply the visual receptive field shifted by $\vec{c}$, as one
might have guessed. Rather, the shape of $F_{A}(\vec{x}\,)$ could be
significantly distorted by the cost function $C_{A}({x})$. At the
moment, the data may be too crude to determine the shape of $F_{A}(\vec{x}\,)$,
but it is still worthwhile to contemplate what sort of shape distortion
may result in our simple model. For illustrative purposes, let us take 
$F_{V}(\vec{x}\,)\propto \exp\left[-(x/l)^2\right]$ to be a Gaussian with $l$
a length scale characteristic of the visual receptive field and 
$C_{A}(x)=\mu {x}^{2}$. Then we predict (with $\vec{c}=(c,0)$ and $\vec{x}=(x,y))$
\begin{equation}\label{FaG}
F_{A}(x,y)\propto \frac{1}{1+\left(x^2+y^2\right)/L^2}
\exp\left[-\frac{(x-c)^2+y^2}{l^2}\right],
\end{equation}
where $L\equiv \sqrt{E_{V}/\mu }$ defines a length scale characteristic of
the relative cost of connecting distant neurons. While there are three length scales 
$L$, $l$, and $c$ inovlved, the shape of $F_{A}(x,y)$ depends only on
the two ratios ${L}/{l}$ and ${c}/{l}$.

\begin{figure}
\begin{center}
\includegraphics[width=8cm]{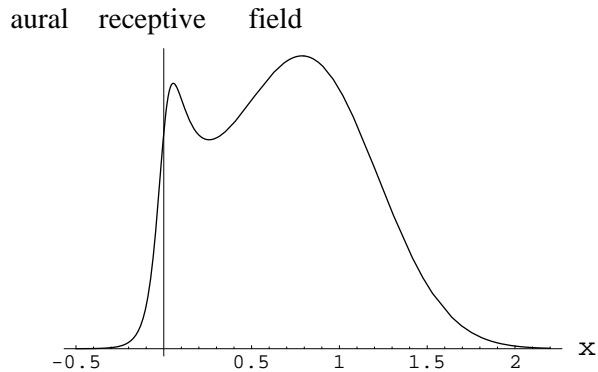}%\\(a)\\
\end{center}
\caption[]{An aural receptive with two peaks, obtained from Eq.~(\ref{FaG}) for 
$l=0.5$, $L=0.1$, and $c=1.1$.}
\label{2peaks}
\end{figure}

We now qualitatively describe the change in the shape of the aural receptive
filed in Eq.~(\ref{FaG}), as the imposed shift $c$ is varied as in the experiments
of Knudsen et al.
(The exact analysis of the extremal points of Eq.~(\ref{FaG}) involves the solution
of a cubic equation which will not be given here.)
Two types of behavior are possible depending on the ratio $l/L$.
For $l\ll L$, where the cost of rewiring is negligible, the function $F_{A}(x,y)$
has a single maximum located at $x\approx c$ (and $y=0$), 
i.e. simply following the imposed shift.
When $l\gg L$, however, 
there is an intermediate range of values of $c$, where the aural receptive
field has two peaks, one close to the origin, $\overline{x}_-\approx cL/l\ll c$,
and another close to $\overline{x}_+\approx c$.
A typical profile with two peaks is depicted in Fig.~\ref{2peaks}.
%For $c\lesssim \ell$, there is a single peak located at $\overline{x}_-\approx cL/l\ll c$,
%while an additional peak appears at $\overline{x}_+\approx c$ for $\ell< c< L$
%(along with a minimum at $\overline{x}\prime\approx Ll/c$).
%A typical profile with two peaks is depicted in Fig.~\ref{2peaks}.
%For $c>l$, the peak at $\overline{x}_-$ disappears, leaving only a single peak 
%at $\overline{x}\approx c$.

\section{Discussions}\label{discussions}

Equation~(\ref{Fa}) is the central result of our study. It provides the optimal linear
filter for a weak signal correlated to a stronger one.
Some specific features of this result in connection with the coupled visual and
aural maps are:
\begin{itemize}
  \item The shape of the aural receptive field is very much controlled by the visual
information, modulated by the costs associated with neural connectivities.
  \item Artificially displacing the two signal sources, as in the case of the prismatic 
spectacles used on the barn owls\cite{BK}, modifies the aural receptive field.
However, the resulting receptive field is not simply shifted (unless the costs
of neural wirings are negligible), but also changes its shape.
  \item Equation~(\ref{Fa}) is the product of two functions, one peaked at the origin
and the other at $\vec x=\vec c$. Depending of the relative strengths and widths
of these two peaks, the receptive field may be more sensitive to signals at the original,
or in the shifted location.
 \item The experiments find, not surprisingly, that adaptation to the prismatic glasses
depends strongly on the age of the individual owl.
This feature can be incorporated in our model with the reasonable assumption that the
cost of neural connections increases with age of the individual.
\end{itemize}

This work is small step towards providing a quantitative framework for deducing the
workings of the brain, starting from the tasks that it has to perform for the organism
to function in its natural habitat.
In this framework, the tasks of the sensory systems are more apparent: to extract the
relevant signals from the background of natural inputs, and as a first step to localize
the source of the signal in the external world.
It is possible to experimentally gather information about the correlations of 
various signals in the natural world,
and there are indeed several studies of the statistics of various aspects of
visual images\cite{natural}.
Of course, such statistics are also specific to the instrument (e.g. camera)
used to obtain the image.
More relevant are psycho-physical studies that probe how individuals parse the
visual information\cite{Malik}.
We are not aware of similar studies on the statistics of natural sounds
in different directions, and their correlations with visual signals.
Such studies may provide part of the material needed for a more detailed study.

The outcome of the procedure outlined in this paper is a set of filter functions, which
are hopefully related to the actual connections between neurons.
The shape and range of such connections can be studied directly by
injection of biocytin dye\cite{dye},
and indirectly by mapping the receptive field of a neuron via a microelectrode probe.
Detailed studies of this kind for the owls reared with prismatic spectacles, and their
comparison with Eq.~(\ref{Fa}) may provide insights about the cost of making
neural connections, another necessary input to our general formalism.

The analytical formalism itself can be extended in several directions.
Already, in Ref.~\cite{BRZ} it was proposed that colored images can be studied
by considering a vector signal $\vec{s}$ ranging over the color wheel.
In regards to different sensory inputs, we may can also ask if and when it is
advantageous to segregate outputs to distinct cortical areas, allowing for distinct
maps $\{O_\nu\}$.
A more ambitious goal is to extend the formalism to time dependent signals,
allowing for filters with appropriate time delays that attempt to take advantage
of temporal patterns in the signals.

\acknowledgments
This work was supported in part by the NSF under grant numbers
DMR-01-18213 (MK), PHY89-04035 and PHY95-07065 (AZ).

%\tighten
%

%\end{references}
%end tighten (references)
\end{document}